*Cover Page:*

# Graphene-Based Nanostructures in Electrocatalytic Oxygen Reduction

Jerzy K. Zak

*Faculty of Chemistry, University of Warsaw, Pasteura 1, PL-02-093 Warsaw, Poland*

Enrico Negro

*Department of Industrial Engineering, Università degliStudi di Padova in Department of Chemical Sciences, Via Marzolo 1, 35131 Padova (PD) Italy*

Iwona A. Rutkowska

*Faculty of Chemistry, University of Warsaw, Pasteura 1, PL-02-093 Warsaw, Poland*

Beata Dembinska

*Faculty of Chemistry, University of Warsaw, Pasteura 1, PL-02-093 Warsaw, Poland*

Vito Di Noto

*Department of Industrial Engineering, Università degliStudi di Padova in Department of Chemical Sciences, Via Marzolo 1, 35131 Padova (PD) Italy*

Pawel J. Kulesza

*Faculty of Chemistry, University of Warsaw, Pasteura 1, PL-02-093 Warsaw,*

emails:   Jerzy.Zak@polsl.pl (JZ)
          enrico.negro@unipd.it (EN)
          ilinek@chem.uw.edu.pl (IAR)
          bbaranowska@chem.uw.edu.pl (BD)
          vito.dinoto@unipd.it (VDN)
          pkulesza@chem.uw.edu.pl (PJK)

**Abstract**

Application of graphene-type materials in electrocatalysis is a topic of growing scientific and technological interest. A tremendous amount of research has been carried out in the field of oxygen electroreduction, particularly with respect to potential applications in the fuel cell research also with use of graphene-type catalytic components. This work addresses fundamental aspects and potential applications of graphene structures in the oxygen reduction electrocatalysis. Special attention will be paid to creation of catalytically active sites by using non-metallic heteroatoms as dopants, formation of hierarchical nanostructured electrocatalysts, their long-term stability, and application as supports for dispersed metals (activating interactions).

# Graphene-Based Nanostructures in Electrocatalytic Oxygen Reduction

**Jerzy K. Zak**, *Faculty of Chemistry, University of Warsaw, Pasteura 1, PL-02-093 Warsaw, Poland*

**Enrico Negro**, *Department of Industrial Engineering, Università degliStudi di Padova in Department of Chemical Sciences, Via Marzolo 1, 35131 Padova (PD) Italy*

**Iwona A. Rutkowska**, *Faculty of Chemistry, University of Warsaw, Pasteura 1, PL-02-093 Warsaw, Poland*

**Beata Dembinska**, *Faculty of Chemistry, University of Warsaw, Pasteura 1, PL-02-093 Warsaw, Poland*

**Vito Di Noto**, *Department of Industrial Engineering, Università degliStudi di Padova in Department of Chemical Sciences, Via Marzolo 1, 35131 Padova (PD) Italy*

**Pawel J. Kulesza**, *Faculty of Chemistry, University of Warsaw, Pasteura 1, PL-02-093 Warsaw, Poland*

*Introduction: fundamentals and applications of the Oxygen Reduction Reaction*

The electrochemically performed oxygen reduction reaction (ORR) leads to a variety of products where the oxygen atoms exhibit the oxidation states -1 (peroxides) and -2 (oxides) [1]. The ORR is characterized by a complex reaction mechanism that is heavily influenced by a large number of factors with a particular reference to the chemical structure and the composition of the interface where it occurs. The ORR mechanism is also strongly affected by the environmental conditions, such as the partial pressure of $O_2$, the temperature and the pH values at the interface. Over the course of the past several decades the ORR has been the subject of extensive investigations, aimed at the elucidation of its rich fundamental features. The ORR is also of enormous practical relevance, as it plays a crucial role in the operation of several families of advanced electrochemical energy conversion and storage devices. The latter include fuel cells, redox flow batteries and metal-air batteries, which are some of the most important actors in the major efforts that are devoted worldwide today to decarbonize the energy sector and fully exploit the potential of renewable energy sources.

The kinetics of the ORR is characterized by large overpotentials (on the order of several hundreds of mV) necessary to achieve a current density that is large enough to be of practical use in any energy conversion/storage application. Such large ORR overpotentials degrade significantly the efficiency of the overall process exploited by the energy conversion/storage device and are one of the most crucial bottlenecks in its operation. For all these reasons, the development of effective electrocatalysts(ECs) able to curtail the ORR overpotential is one of the most crucial and hot topics in the modern research on advanced electrochemical energy conversion and storage devices.

The ORR may occur by means of two main pathways: (i) a *"direct"* four-electron process, yielding products where oxygen is in the -2 oxidation state (*e.g.*, water or oxides, depending on the chemical environment); and (ii) an *"indirect"* two-electron process, giving rise to peroxide species (*e.g.*, hydrogen peroxide). The latter may subsequently undergo an additional reduction step and yield the same products obtained in pathway (i). A high ORR selectivity in pathway (i) is preferred for applications in electrochemical energy conversion and storage devices. Indeed, pathway (i) does not generate oxidizing species (*i.e.*, peroxides) that may degrade the functional materials involved in the operation of the system and shorten its operating lifetime. In rough terms, the ORR occurs through three main successive steps, as follows: (i) the $O_2$ reactant approaches the interface where the ORR is taking place; (ii) electrons are transferred to oxygen; and (iii) oxygen-based products are removed from the interface. The fastest overall ORR kinetics is achieved on active sites based on platinum, that is able to strike the best compromise between the speeds of the various reaction steps. As the pH is raised, the ORR kinetics of Pt-based active sites is slightly decreased. This phenomenon is associated with a more facile clogging of active sites by oxygen-based species. On the contrary, the increase in pH significantly improves the ORR kinetics on most other active sites that do not include platinum or other platinum-group metals (PGMs). This outcome is typically interpreted admitting that a higher pH facilitates ORR Step (ii), *i.e.*, the transfer of electrons to oxygen.

Several techniques are available to evaluate the ORR performance and reaction pathway; the most accurate results are obtained using the linear sweep voltammetry (LSV) at the thin-film covered rotating disk electrode (RDD), when current *vs.* potential curves are recorded at different rotating speeds. This technique allows to neglect the issues associated to limitations of mass transport phenomena, focusing only on the ORR kinetics. The selectivity in the *"direct"* pathway can also be determined accurately, especially if a rotating ring-disk electrode (RRDE) is available. Indeed, the latter device is able to detect directly the peroxide intermediates.

*Graphene: Fundamentals and potential applications in ORR electrocatalysis*

Graphene is recently-discovered allotrope of carbon that ideally consists of bidimensional, one-atom-thick layers. Graphene exhibits outstanding physicochemical properties, including: (i) an extremely large specific surface area (up to ca. 2600 $m^2 \cdot g^{-1}$); (ii) a very high mobility of the charge carriers (up to more than 200000 $cm^2 \cdot V^{-1} \cdot s^{-1}$). These features set graphene far apart from conventional graphitic carbon, that is characterized by significantly lower specific surface area (on the order of about 250 $m^2 \cdot g^{-1}$), mobility of charge carriers (about 6000 $cm^2 \cdot V^{-1} \cdot s^{-1}$). These features make graphene a very promising candidate for the development of electrocatalysts (ECs). In particular, the very large specific surface area and the lack of cavities and pores would enable an outstanding dispersion of the active sites and a facile mass transport of reactants and products. Furthermore, the high mobility of charge carriers would prevent the occurrence of ohmic drops. Pristine graphene ideally includes only carbon atoms exhibiting a $sp^2$ hybridization and strongly bound with one another through covalent bonds; furthermore, both faces of the graphene sheet are covered by a symmetrical and uniform distribution of π electrons characterized by a very low polarizability. These features cause the *"ideal"* graphene to exhibit a high chemical inertia. On one hand this characteristic is advantageous for an EC, in the sense that *"ideal"* graphene is expected to exhibit a high tolerance towards degradation upon electrochemical operation, resulting in a high durability. On the other hand, too low reactivity effectively inhibits all the interactions with the species involved in the electrochemical process of interest (in our case, the ORR), resulting in a poor performance. Therefore this last issue needs to be addressed when improving the reactivity of pristine graphene by introducing on each graphene layer a suitable surface concentration of active sites with a well-controlled chemical composition and structure. Ideally, this would result in a high ORR performance and still afford a durability level compatible with the particular application at hand. This objective can be approached by at least two avenues, that can be classified as follows:

- Graphene can be used to support directly suitable nanoparticles, that bear the desired active sites. The ORR performance is maximized by modulating the chemical composition and the morphology of the supported nanoparticles in the resulting nanocomposite.
- Active sites can be doped into the graphene layer, that undergoes chemical modification. One or more heteroatoms (*e.g.*, N, B, S) are introduced into the graphene layers; the best performance in the ORR is achieved if the heteroatoms coordinate additional metal species.

*Graphene-Metal Specific Interactions*

Strong effect of structural defects concentration within graphene quantum dots on oxygen reduction performance of platinum nanoparticles was postulated [3]. It was shown (by FTIR, Raman and XPS analyses) that the hydrothermal treatment at higher temperatures and longer times partially removes oxygen functionalities and leads to the increase of the $sp^2$ carbon domains. The dependence of the electrocatalytic activity as function of the defects concentration showed that, in the optimum catalysts the binding energies of Pt 4f were the lowest thus implying minimum charge transfer from Pt to graphene quantum dots.

Such interactions were also observed experimentally (and confirmed theoretically by density functional theory calculations) in the case of platinum nanoparticles deposited onto N-doped reduced graphene oxide [4]. XPS experiments showed that about 25% of Pt nanocenters were chemically bound to nitrogen atoms which allowed for an electronic transfer from platinum to carbon. The increased activity towards oxygen electroreduction (when compared to other substrates such as reduced graphene oxide, graphene oxide and Vulcan XC-72) was attributed to the elongation of O-O distance, thus facilitating the $O_2$ dissociation, although simultaneously the weaker $O_2$ adsorption was observed.

More recently highly dispersed and very active $Pt_2Pd$ alloy (obtained from platinum and palladium phtalocyanines) has been embedded in nitrogen-rich graphene nanopores [5]. The XPS analysis showed that $Pt_2Pd$ alloy has exhibited strong interactions with graphene nanopores through nitrogen atoms. DFT calculations revealed that the adsorption of the alloy was much stronger on nitrogen-rich graphene nanopores when compared to graphene and nitrogen-doped graphene supports implying that the stability of the catalyst is most influenced by nanopores (and to less extent by nitrogen). As a result, agglomeration of the nanoparticles has been inhibited. Furthrmore, the d-band center of Pt and Pd has been down-shifted thus indicating the modulation of $Pt_2Pd$ electronic properties and facilitating fast electron transfers.

*Graphene structures with non-metallic heteroatoms as dopants*

Different routes for graphene or graphene oxide (GO) are currently explored based on their existing active sites; these reported in the review assume the creation of catalytically active sites by doping using elements, which affect the existing graphene electronic structure as donors or acceptors.

The former era of intensive exploration of carbon nanotubes (CNT) led to finding that vertically aligned CNT (VA CNT) containing nitrogen may effectively catalyze ORR with a better long term stability than commercially available Pt-based electrodes. The incorporation of electron-accepting nitrogen atoms into the conjugated nanotube carbon plane was found effective in creation of a relatively high positive charge density on the adjacent carbon atoms [6].The discovery of graphene, which is, as many authors say 'the mother of all graphitic forms' has opened a new research perspective since this 2D material of high aspect ratio (lateral size to thickness) along with its good mechanical properties, rich electronic states, excellent 2D conditions for electron transport, and large specific surface area may create a new type of electrode along with usually carbonic support. The most often cited paper [7] describes the CVD (chemical vapor deposition) procedure in which ammonia together with methane/hydrogen gas mixture was used to form N-doped graphene film on $Si/SiO_2$ wafers with Ni layer as catalyst. The careful procedure led to absence of any Ni residue as confirmed by XPS spectrum. This spectrum at high resolution in the region of N 1s revealed a presence of both, pyridine-like and pyrrolic nitrogen atoms in the graphene structures, similarly as found earlier in VA CNT [6]. The N-atoms have been incorporated into the graphene hexagon rings at estimated N/C atomic ratio of *ca.* 4%. The electrochemical measurements (RRDE) in alkali solution for the film on GC electrode for both, the N-doped and undoped graphene, and for Pt-loaded carbon (Vulcan XC-72) (Pt-C) indicate on two-step (-045 and -0.7 V), two-electron processes on pure graphene electrode, whereas the process on N-doped film was one-step (-0.25 V), 4-electron ORR. Although the onset potential for Pt-C electrode was about 100 mV more positive, the steady-state current density was ca. 3 times higher for the N-doped graphene electrode over large potential range. Other measurements, including exposition to methanol or CO have also demonstrated that a poison effect and long-term stability are also better for N-doped graphene than for Pt-C electrode. Earlier computational studies were predicting ability of carbon materials with N-dopants at specific sites to exhibit catalytic activity in systems like fuel cells. Several simulations were performed [8] for models composed of graphene sheets doped with N-atoms at different positions. Free energy profiles were demonstrated for adsorption of $O_2$ molecule in terms of distance between its center position and the target adsorption site. The calculations suggested that carbon type of catalyst displays a potential ORR catalytic activity that is enhanced at specific graphene sites like zigzag edges, if N- atom is located there. Then the subsequent reaction preferentially takes that path, which leads to formation of two water molecules with lower free-energy barrier.

The catalyst-free thermal annealing process was proposed [9] for a large scale synthesis of N-doped graphene based on low-cost industrial material melamine as nitrogen source. In this process graphene oxide (GO) synthesized by commonly applied Hummers' method was mixed with melamine and then grinded. The thermal processing forms first carbon nitride, which in turn is decomposed to nitrogen, making possible its attack on these carbon sites, which come from decomposition of GO. Selection of conditions for the process and the melamine to GO ratio let to manipulate with the doping level of the final product, even up to 10.1 %. The high resolution XPS analysis allowed to distinguish three types of nitrogen in the product, beside the pyridinic- (dominating) and pyrollic-, a graphitic-N atom was also assigned to the resulting spectra. The recorded CV curves indicated on the catalytic effect of nearly 0.1 V for all doping levels as compared to bare GC and GC/graphene (undoped) electrode, besides, the reduction current observed in alkali solution was increasing with the doping level.

Another approach was demonstrated in fabrication of graphene-based carbon nitride (G-CN) nanosheets with individual dispersion of the graphene layers [10]. Employment of graphene-based mesoporous silica nanosheets as a template and ethlenediamine and carbon tetrachloride as CN precursor led to G-CN nanosheets with a high nitrogen content, large surface area (542 $m^2g^{-1}$) and enhanced electrical conductivity. This is in contrast to carbon nitride (CN), in form of nanosheets prepared also using the same template, which due to a higher nitrogen content than in case of G-CN should be an excellent catalyst for ORR, but its poor electrical conductivity ($< 10^{-2}$ $Scm^{-1}$) is a major obstacle for its applications in fuel cells. The procedure led to three different types of electrode material with differentiated ratio of pyridinic- *vs.* pyrollic-N at approximately the same content of graphitic-N in each case. However, since the total amount of N-atoms in these three materials was changing from ca.

6% to 20%, the ratio of pyridinic to pyrollic N-atoms cannot be treated as the only factor affecting the observed catalytic activity. The LSV at RDE data indicate for the first time that both electrical conductivity and the content of pyridinic nitrogen are the vital factors for achieving a highly effective catalyst of ORR. This is in agreement with the earlier quantum mechanical calculations [8], and other experimental investigations reported in the references. That catalyst is also unaffected by methanol additions. The same group extended the studies with graphene-oxide mesoporous silica template toward S-doped graphene [11]. Their universal procedure with silica template applied the annealing in gas atmosphere of $NH_3$ or $H_2S$ to introduce heteroatom into GO structure. The numerous mesopores formed at GO sheets with the size of about 2 nm led to the BET surface area of 1051 $m^2g^{-1}$ at the uniform thickness of 15 nm. The procedure of the high temperature annealing in the gas atmosphere of $NH_3$ or $H_2S$ has also a strong effect on reduction of GO. XPS data revealed that spectrum of GO is nearly totally eliminated above 500 $C^0$. The tested catalysts were prepared at temperatures ranging from 600 to 1000 $C^0$ for NG catalyst and of 100 degree lower in case of SG catalyst. The electrochemical activity given as the kinetic limiting current density at – 0.50 V was in case of both catalysts (obtained at the highest temperature) comparable, like in case of SG or even higher, the case of NG, than that obtained for commercial Pt-C catalyst. The studies have also demonstrated for the first time that sulfur can be doped into graphene, dominantly as thiophene-like S, whereas in case of N-heteroatom three forms are present as described above. All the data reported by the Authors gave a strong evidence that both N and S-doped graphene sheets show excellent electrocatalytic activity, long durability and high selectivity as catalyst for ORR. They also predict that the same synthetic approach can be applied to synthesize a series of heteroatom-doped graphenes with B, P, and F.

The dual N and S doped graphene, N-S-G, is also reported as an excellent catalyst for ORR, and significantly better than graphene doped solely with S atoms (S-G) or with N atoms (N-G) [13]. The outstanding performance of that system is explained by dual activation of graphene carbons, which is supported by both experimental results and quantum chemistry calculations. In the synthesis of the catalyst colloidal silica nanoparticles (12 nm) were used as a structural template to create sites of large mesopores. GO was the initial material, melamine and benzyl disulfide were used as N and S precursors. The doping process occurs at $900^0C$ in Ar atmosphere. Electrochemical data revealed that N-S-G displayed a high ORR onset potential of -0.06 V, close to (30 mV) Pt/C electrode and more positive than that of N-G or S-G (ca. – 0.18 V) as recorded by LSV on RDE (all potentials *vs.* Ag/AgCl reference electrode). The synergetic effect of two dopants in ORR was confirmed by DFT calculations.

Whether an atom creates an active sites depends on the charge density and spin density (dominating). For N-G the doping occurs from more electronegative atom, in case of S-G the electronegativity of C and S are comparable. The catalytic activity comes from mismatch of the outermost orbitals of both dopants, so the resulting positive charge is located at S atom, which is a catalytic center for ORR. In conclusion, the synergetic effect results from redistribution of spin and charge densities, which in case of dual doping creates a large number of carbon atom active sites.

Based on previous studies on VA CNT co-doped with B and N in which a synergetic effect was observed for VA BCN as catalyst of ORR [12a] the studies were undertaken on that concept, but based on graphene instead of VA CNT [12]. Since the CVD synthesis is practically too complicated and too expensive for mass production, a simpler procedure needs to be applied in the catalyst production. The choice was a development of thermal annealing of GO in the presence of boric acid and ammonia which allows to tune the B/N doping level. Also, the first-principles calculations were performed to support the results indicating on the superior catalytic properties of BCN graphene. A wide characterization of the synthesized BCN graphene samples was provided which includes XPS, FTIR, and Raman spectra for all the proportions between the carbon and dopants. The electrochemical data demonstrated the LSV curves of ORR on BCN graphene with different compositions; they were compared with the commercial Pt/C catalyst. The latter confirms an excellent electrocatalytic properties for nearly equally doped sample $B_{12}C_{77}N_{11}$, which appears to be a better catalyst than Pt/C. This electroactivity was also studied by density functional calculations (DFT). Several BCN graphene models were designed, comparable with these obtained experimentally. The calculated HOMO-LUMO values were then used for estimation of reactivity, since a small energy gap implies a low kinetic stability and a high reactivity. As compared to the pure graphene, the substitution of C by B and N leads to a smaller energy gap. The detailed analysis of the HOMO–LUMO data involves also the conductivity of the material. As found from the analysis, the experimental observation are in agreement with theoretical calculations. As the Authors conclude, the thermal annealing graphene oxide in presence of boric acid under ammonia can provide a simple but efficient and versatile approache to low-cost mass production of BCN graphene as an efficient metal-free ORR electrocatalyst for fuel cells. Important finding based on detailed studies using DFT calculations are presented for the model N-doped graphene [14a]. At the experimental conditions taking into account solvent, surface adsorbates, and coverages two mechanisms of the catalytic reaction are considered, dissociative and associative. In both cases the first step is analogous, in which molecular oxygen is bond to surface active site. Next step however involves addition of proton to individual surface *O atoms (after dissociation) or to surface molecule *$O_2$ (associative case). It is postulated that N-graphene surface is covered with O by 1/6 of its monolayer. The calculated detailed energy profiles are presented for the series of consecutive reaction steps in alkali solution, from oxygen molecule in water to 4 $OH^-$. Each step with its energetic barrier allows to conclude that the associative, $4e^-$ route is favored. Several conclusions were drown from this important work, among them (i) the water effect is essential in that mechanism, without water $O_2$ molecule cannot adsorb on N-graphene surface, (ii) the dissociation barrier for *$O_2$ is too high, therefore the associative mechanism is more likely, (iii) the removal of adsorbed *O from N-graphene determines the rate of ORR process. Similar, important theoretical studies on these two types of mechanisms were presented for the B-doped graphene [14b], considered both in Langmuir-Hinschelwood and Eley-Rideal schemes. Here, the Authors found that among all intermediates and transitions considered, the most significant species is that formed from adsorption of molecular oxygen on positively charged B-doped atom. This C-B-O-O open shell intermediate cannot be conceived in case of pure graphene and was not found on N-doped graphene. Also, the pH effect on the ORR is widely discussed in the proposed model in terms of the electrode overpotential vs. pH.

Another approach, in which N and B were sequentially incorporated into selected domain of graphene demonstrated an enhanced synergistic coupling effect in the electrocatalytic ORR [15]. That new method proposed, also prevents a formation of inactive by-products. B,N-graphene was prepared from solution exfoliated GO in two steps: the first was annealing in $NH_3$ atmosphere, then B was introduced from $H_3BO_3$ precursor at elevated temperature. For the obtained catalyst the electron-transfer number was determined as 3.97 in ORR, which was very close to 3.98 obtained at the same conditions for the commercial Pt-C. Also, the kinetic limiting current was comparable for these two systems, 13.87 $mAcm^{-2}$ for B,N-graphene vs. 14. 64 $mAcm^{-2}$ for Pt-C. The Authors emphasize an excellent methanol tolerance of that catalyst. The experimental results are supported by the results of DFT calculations.

A three-dimensional nitrogen doped catalyst was designed, in which a composite was prepared from carbon nanotubes and graphene NCNT/G [16]. This structure was obtained in the process of pyridine pyrolysis over a graphene-sheet supported Ni catalyst. The tangled NCNT of several hundreds of nm form a quasi-aligned arrays with a nitrogen content of 6%. As shown, the electrochemical performance of this catalyst (onset potential of − 0.20 V, the electron transferred number of 3.5) is located between the commercial Pt/C catalyst and undoped CNT/G system.

It has been demonstrated in many papers that the doping heteroatom is located at the active edges of graphene framework. If the edges are produced during the doping process then the procedure may be especially effective. That idea was proposed in the paper [17] describing a large-scale production of edge-selectively functionalized graphene nanoplatelets (EFGnPs) *via* ball milling in the presence of (respectively) hydrogen, carbon dioxide, sulfur trioxide, or carbon dioxide/sulfur trioxide mixture. When the products were then exposed to air moisture the resulting graphene edges were functionalized with hydrogen (HGnP), carboxylic acid (CGnP), sulfonic acid (SGnP) or carboxylic/sulfonic acid (CSGnP). All these modified graphene products were then used as catalysts in ORR, which let to propose the series of catalytically active modified graphenes in their increasing activity SGnP>CSGnP>CGnP>HGnP>pristine graphite. Earlier the same Authors [17a] have demonstrated that the edge-carboxylated graphene without basal plane distortion can be obtained via simple milling of graphite in the presence of dry ice. The products of these processes are highly dispersible in polar solvents. The procedure allows to investigate a pure edge effect, practically without any heteroatom-doping on the basal plane. Generally, the functionalization of graphite edges together with doping process creates an interesting possibilities for many catalytic processes.

A dose of skepticism was introduced to the area of development of metal-free catalysis when the paper entitled ""Metal-Free" Catalytic Oxygen Reduction Reaction on Heteroatom-Doped Graphene is Caused by Trace Metal Impurities was published"[18]. This strong statement was supported with several experiments, which gave a proof of presence of some metal impurities in all samples of graphene oxide prepared according to common procedures proposed by Hummers or Staudenmaier protocol. The list of impurities determined using ICP-MS method consists of Fe, Co, Ni, and Mn, whereas the concentration range is between single ppm up to ca. 8000 ppm. LSV data are also presented, which demonstrate effects of these impurities on ORR. The strongest effect comes from $MnO_2$, which might be found in samples of graphene prepared by Hummers method even at the level of 8000 ppm. When carefully reading most of papers in the area, having in mind the message of that particular paper, one may easy find that most of authors take care about the problem of metal contents, besides, it has been always demonstrated a kind of background signal recorded for reference sample, which is an undoped material. More, when an effect of a dopant is demonstrated, it comes usually as an effect of variable concentration of a given dopant. Therefore, this paper should be treated as an important warning, not a voice questioning earlier, high quality research presented in most of papers, at least in these presented in that review.

*Hierarchical Nanostructured Electrocatalysts*

A material exhibits a *"hierarchical"* structure if its features can be broken down into at least two different size levels. Under this definition, most nanocomposite ECs including graphene and related materials (GRMs) could be considered hierarchical. Indeed, such systems comprise both:

- A first component with a *"higher"* size level, on the order of a few hundreds of nanometers or more (the single GRM sheet);
- A second component with a *"lower"* size level, on the order of a few tens of nanometers or less (the nanoparticle bearing the ORR active sites).

However, the common usage of the term *"hierarchical"* in the modern literature on GRM-based ECs is somewhat different. A *"hierarchical"* GRM-based EC must include a nanostructured support, where the GRM sheets are *"spaced"* by another component with a different size level that is typically not directly involved in the electrocatalytic function under study. The most common examples of *"hierarchical"* GRM-based ECs includes systems where GRM sheets are spaced by other components, including carbon nanotubes, carbon nanoparticles, other inorganic nanostructures (*e.g.*, based on inert oxides), and various coatings (*e.g.*, macromolecules).

A very serious issue that is observed in conventional GRM-based ECs is the restacking of GRM layers in the final material at the end of the synthesis process. This phenomenon is prompted by the strong van der Waals interactions between the surfaces of the GRM sheets. In the resulting *"restacked"* system, a large fraction of the active sites is buried between GRM layers; consequently, the mass transport of reactants and products is seriously hindered. Another drawback suffered by *"restacked"* GRM-based ECs is concerned with the transport of electrons. The latter can occur in the *"through-plane"* direction, or by *"hopping"* between different GRM layers and thus suffer from a poor conductivity [19], much inferior in comparison with the outstanding values associated with pristine, single-monolayer graphene in the *"in-plane"* direction. Both of the issues outlined above become more relevant as the current density is raised; accordingly, their impact in the operation of compact, high-J applications (*e.g.*, PEMFCs) is crucial.

Typical hierarchical GRM-based ECs exhibit a complex 3D morphology, that was obtained by adopting a variety of synthetic approaches [20]. It starts with graphene oxide (GO) which is dispersed in water together with pyrrole and iron acetate. The resulting product undergoes a hydrothermal treatment, that gives rise to self-assembly; the product is freeze-dried and eventually undergoes a pyrolysis process, yielding the final hierarchical GRM-based EC (in this case, a hybrid aerogel. This preparation process includes several features that are commonly found in the literature, and is an excellent case study. The authors use a GO precursor, that is easily suspended in water; accordingly, restacking does not occur. Iron acetate provides the iron atoms to be embedded in the ORR active sites; pyrrole acts a binder, to stabilize the 3D morphology of the hierarchical EC and to facilitate the anchoring between the active sites and the GO support through N atoms. The hydrothermal treatment triggers the formation of the hierarchical 3D morphology, that is eventually consolidated by the pyrolysis process. The latter also plays a crucial role to reduce the GO (that is known to exhibit a very poor electrical conductivity) to graphene-like sheets. The resulting EC was able to demonstrate a very good ORR performance in an alkaline environment, both in terms of ORR onset potential and selectivity in the 4-electron pathway. In another case [21], an aqueous suspension of GO, Pt and Ru precursors was electrochemically reduced on a matrix of $TiO_2$ nanotubes ($TiO_2$-NTs). It was clearly demonstrated that the ORR performance of the proposed EC in an alkaline medium is significantly improved upon the introduction of the rGO in the morphology of the hierarchical support, whose porosity (that is bestowed by $TiO_2$-NTs) is expected to allow for a facile transport of reactants and products.

A GRM-based EC with a hierarchical morphology can also be obtained by self-assembly of different 2D systems [22]. A suspension of graphitic carbon nitride nanosheets (g-$C_3N_4$) was prepared by exfoliating the product of the pyrolysis of a melamine sample; a suspension of GO was added and the resulting self-assembly process gave rise to a hierarchical 3D porous g-$C_3N_4$ architecture, which finally undergoes a photoreduction process to give the final EC. The resulting EC exhibited a large surface area and a hierarchical porous structure which promotes the diffusion of $O_2$, thus improving the ORR performance. It was also demonstrated that, with respect to its pristine g-$C_3N_4$ EC analogue, the hierarchical 3D porous g-$C_3N_4$ also apparently exhibits a much improved selectivity in the 4e- ORR pathway. Another example of a GRM-based EC with a hierarchical morphology is reported in [23]. The preparation is carried out by a hydrothermal process at low temperature (T = 180°C) of a precursor consisting of GO, oxidized multiwalled carbon nanotubes (OCNT) and ammonia. The resulting hierarchical GRM-based EC, labeled "NG-NCNT", exhibits a highly porous structure including reduced graphene oxide (rGO) and carbon nanotubes; both components are doped with nitrogen. This study is of particular interest since the final hierarchical GRM-based EC is obtained using very simple equipment, and without the need to apply a high-temperature pyrolysis process step. Accordingly, the process is easy to scale-up and does not require large amounts of energy. It is further demonstrated that the resulting hierarchical GRM-based EC exhibits a remarkable performance in the ORR, that is obtained by the synergic interaction between the two components obtained following the preparation process. Indeed, it is readily evident that the nitrogen-doped rGO and the nitrogen-doped multiwalled carbon nanotubes exhibit a significantly inferior ORR performance as they are taken on their own or simply mixed together. Finally, the proposed hierarchical GRM-base EC exhibits a high selectivity in the 4-electron pathway.

A further interesting example of a hierarchical GRM-based EC is proposed in [24]. In this case, the hierarchical support is obtained by impregnating a polyurethane foam with a water suspension including GO sheets and urea. The product is dried and calcinated at 900°C in an Ar atmosphere; in a second step, the hierarchical support is loaded with Pt nanoparticles by means of a conventional microwave-assisted polyol process, yielding the final Pt/3D-NG product. This study is particularly interesting as one component of the hierarchical support is actually a system exhibiting macroscopic pores (*i.e.*, the polyurethane form); in this case, the GO sheets decorate the foam, providing the attaching sites for the Pt nanoparticles bearing the active sites. The authors demonstrate that this approach is promising to obtain high-performing ECs that are able to promote effectively both the methanol oxidation reaction and the ORR. It was also demonstrated that the introduction of N in the hierarchical support plays a crucial role to improve the durability of the proposed hierarchical GRM-based EC upon accelerated ageing, as the Pt-N interactions inhibit the agglomeration of the Pt nanoparticles bearing the active sites and the corresponding loss of electrochemically-active surface area (ECSA).

A very innovative and extremely flexible approach for the preparation of hierarchical GRM-based ECs consists in covering a GRM *"core"* with a carbon nitride (CN) *"shell"*, that stabilizes the metal alloy nanoparticles bearing the ORR active sites in *"nitrogen coordination nests"*[25]. These materials are obtained by impregnating the GRM *"core"* with a suitable hybrid inorganic-organic precursor; the product then undergoes a pyrolysis process to yield the final hierarchical GRM-based ECs. The products include a very disordered stacking of graphene sheets; the CN *"shell"* is very porous, and embeds well-dispersed $PtNi_x$ alloy nanoparticles. The ORR performance as determined by LSV-TF-RRDE is comparable with that of the Pt/C reference; an improved selectivity in the 4e-pathway is detected.

*Other electrocatalytic applications*

The graphene-based materials can affect reactivity of electrocatalysts [26,27]. Due to large surface area, unique electronic properties and a high physico-chemical stability, graphene-type materials were often used for dispersing noble metal nanoparticles [28,29]. For example the graphene-supported Pd nanocrystals exhibited appreciable catalytic activity during the formic acid electrooxidation [30]. This observation was particularly sound when using GO as the template for single-crystalline Pd square nanoplates encased within [100] facets [31].

An interesting case is the oxidation of alcohols (methanol, ethanol, ethylene glycol and glycerol) at platinum-ceria/graphene nanosheet (Pt-CeO$_{2-x}$/GNS) catalysts in 1 mol dm$^{-3}$ KOH solution [32]. The higher current densities were attributed to existence of the abundant oxygen-containing species and possibility of removal of the poisoning reaction intermediates. Graphene nanostructures were believed to enhance interactions between noble metal centers and metal oxide sites.

In another example, the graphene-supported metal dimers [33], e.g. using a Cu dimer doped into graphene with adjacent single vacancies, were considered for the electroreduction of CO$_2$ toward CO production [34]. Furthermore, the competing hydrogen evolution reaction was effectively suppressed by the Cu-terminated armchair graphene nanostructures. It was postulated that the edge-decoration f graphene nanoribbons offered great flexibility for tuning the catalytic efficiency and selectivity during CO$_2$ electroreduction [35-38].

Graphene being a monolayer of carbon atoms arranged in a two-dimensional honeycomb network can offer substantial benefits with regard to mass transfer and charge transport, by providing shorter effective lengths for both ionic and electronic transport [39-42]. On the other hand, palladium (Pd) has been reported to possess excellent properties to facilitate electrocatalytic reduction due to its superior ability to form surface-adsorbed atomic H, a highly activated intermediate hydrogen radical [42]. The Pd-reduced graphene oxide (rGO) modified granular activated carbon (Pd-rGO/GAC) system exhibited particularly high electrocatalytic activity toward reduction of bromate, a model highly-inert probe for catalytic electroreductions. With the unique electronic structure of rGO sheets, the electroreduction of

$H_2O$ to atomic H on the Pd particles can be significantly accelerated, leading to a faster reaction rate of $BrO_3^-$ with atomic H [42,43].

**Evaluation and benchmarking of the ORR catalysts' performance and durability**

The ultimate test to ascertain the effectiveness of ORR EC for application in electrochemical energy conversion and storage devices would require implementation in prototypes of practical systems operating under conditions for both performance and durability. This approach is often difficult to pursue for both fundamental and practical reasons. On the fundamental level, the performance and durability of electrochemical energy conversion and storage device are determined by several other factors beyond those associated with the ORR electrocatalyst, e.g. such as the electrolyte conductivity, the compatibility between all the different functional materials involved in fabrication of devices and their tolerance to ageing. Durability tests, which are crucial to gauge the feasibility of the material in real applications, are also very time-consuming. Voltammetric techniques including rotating ring-disk voltammetry are promising approaches in this respect. Although a ring-disk electrode for the direct detection of peroxide intermediates are widely adopted in the literature. There is no agreement, however, on the type of reference electrode to be used in the experimental setup. Careful utilization of RHE reference is critical, since even a small shift in the potential would affect strongly the absolute value of ORR performance, especially at the lowest ORR overpotentials where the faradic currents are low. Furthermore, the contribution of capacitive currents must always be removed, to isolate the ORR faradic currents.

Evaluation of the ORR performance and durability is especially important for GRM-based ECs, both conventional and exhibiting a hierarchical morphology, that typically exhibit a very large specific surface area. The ohmic drop correction must also be executed, to filter out the effects associated to conductivity of the electrolyte and geometry of the electrochemical setup. Finally, the contributions ascribed to the transport of oxygen through the diffusion layer close to the electrode surface are to be removed. All these steps, if carried out properly, allow to determine unambiguously and precisely the kinetic currents associated with the ORR process as a function of the overpotential for each EC. The currents should be typically normalized against the mass of the precious metal deposited on the electrode surface, or ECSA characteristic of the precious metal (as determined by techniques such as hydrogen adsorption/desorption, or CO stripping). It is also reasonable to normalize faradic currents versus the overall surface area of the EC deposited on the electrode. The values of specific surface area can be determined by nitrogen sorption techniques or by evaluation of the double-layer capacitance.

There are two main approaches to compare kinetic behavior of different ECs toward the ORR. The first one (useful for Pt- and Pd-based catalysts) is based on the selection of a potential (typically 0.9 V *vs.* RHE), and the normalized faradic currents characteristic of different ECs are compared. The main limitation of such an approach is inability to compare ECs largely differing in their performance levels. The second approach requires pre-selection of the current density at which the systems' potentials are determined and compared. When concentrating on the so called onset ORR potentials developed at low current densities, e.g. low enough to yield currents on the order of 1/20 of the diffusion-limited faradic current. Under such conditions, the corrections due to the diffusion limitations are very small and can be neglected. Consequently, comparison of ECs (e.g. noble-metal-free systems) exhibiting very different performance levels is feasible.

Benchmarking of the ORR performance of ECs is generally carried out with use of the conventional Pt/C reference known as the cathode material of PEMFCs. For the representative Pt/C reference system, the mass activity at 0.9 V (*vs.* RHE) is equal to 0.12-0.25 A/mg$_{Pt}$; these values could be correlated with the onset potentials of 0.95-0.96 V (*vs.* RHE). But, as yet, the reference loadings for various ECs have not been standardized. This fact becomes an issue when performances of the Pt/C reference and Pt-free ECs are compared. When ECs are on the order of 100-1000 µg/cm$^2$ (namely much higher than that adopted for the Pt/C reference), mass transport could become a limiting factor. Furthermore, there are no commonly accepted levels for the ORR performance levels for electrochemical energy conversion and storage systems other than fuel cells (*e.g.*, for Li-O$_2$ batteries). Among the final important issues are the long-term stability and the choice of a proper electrolyte with the desired pH, depending on intended application. It is expected that low-temperature fuel cells (PEMFCs) would be able to operate for 5000-40000 h, depending on the application (automotive *vs.* stationary). Although there is no consensus about practical durability but it is expected that, at the least, the loss in ORR overpotential after 10000 potential cycles should be no more than 30-40 mV.

*Summary*

Development of advanced supports comprising GRMs and exhibiting a hierarchical morphology is a very promising avenue to fully exploit the unique properties of graphene and its derivatives. It was clearly demonstrated that this approach can be used for all the different families of graphene-based ORR ECs, including both systems bearing precious metal nanoparticles (*e.g.*, Pt) and all the different types of *"Pt-free"* materials. A good control of the morphology, composition and structure of the hierarchical ECs can be achieved. Despite these successes, there are still a number of issues that must be addressed by the research in this area. In particular, while it is often possible to gauge the impact of the hierarchical morphology on the overall ORR performance and reaction pathway, a clear understanding of the fundamental mechanisms underlying the achievement of such behavior in the ORR is still missing. In particular, it is often unclear how the interactions between the active sites, the GRM sheets and the other component(s) of the hierarchical support actually affect the ORR kinetics and mechanism. Finally, more efforts must be devoted to study how the ORR performance of the hierarchical GRM-based EC is transferred from the typical *"ex-situ"* electrochemical studies to practical devices, also emphasizing more the issues associated with ageing. This objective can only be reached by implementing the proposed hierarchical GRM-based ECs in operating devices, to be tested in operating conditions for performance and durability.

**Acknowledgements**

We acknowledge the European Commission through the Graphene Flagship – Core 1 project [Grant number GA-696656] and Maestro Project [2012/04/A/ST4/00287 (National Science Center, Poland)].

*References*

1. Shao, M., Chang, Q., Dodelet, J. P., Chenitz, R.(2016) Recent Advances in Electrocatalysis for Oxygen Reduction Reaction, Chem. Rev. **116**, 3594
2. O'Hayre, R., Cha, S. W., Colella, W., Printz, F. B. (2006) Fuel Cell Fundamentals, Wiley, New York.
3. Song, Y., Chen, S.W. (2014) Graphene quantum-dot-supported platinum nanoparticles: defect-mediated electrocatalytic activity in oxygen reduction. *ACS Applied Material Interfaces* **6**, 14050–14060.
4. Ma, J., Habrioux, A., Luo, Y., Ramos-Sanchez, G., Calvillo, L., Granozzi, G., Balbuena, P.B., Alonso-Vante, N. (2015) Electronic interaction between platinum nanoparticles and nitrogen-doped reduced graphene oxide: effect on the oxygen reduction reaction. *Journal of Materials Chemistry A* **3**, 11891–11904.
5. Zhong, X., Qin, Y., Chen, X., Xu, W., Zhuang, G., Li, X., Wang, J. (2017) PtPd alloy embedded in nitrogen-rich graphene nanopores: High-performance bifunctional electrocatalysts for hydrogen evolution and oxygen reduction. *Carbon* **114**, 740-748.
6. Gong, K, Du, F., Xia, Z., Dustock, M., Dai, L. (2009) Nitrogen-Doped Carbon Nanotube Arrays with High Electrocatalytic Activity for Oxygen Reduction. *Science* **323**, 760-764.
7. Qu, L., Liu, Y., Baek, J. B., Dai, L. (2010) Nitrogen-Doped Graphene as Efficient Metal-Free Electrocatalyst for Oxygen Reduction in Fuel Cells. *ACS Nano* **4(3)**, 1321-1326.
8. Ikeda, T., Boero, M., Huang, S. F., Terakura, K., Oshima, M., Ozaki, J. (2008) Carbon alloy catalysts: active sites for oxygen reduction reaction. *Journal of Physical Chemistry C* **112(38)**, 14706-14709.
9. Sheng, Z. H., Shao, L., Chen, J.J., Bao, W. J., Wang, F.-B., Xia, X.-H. (2011) Catalyst-Free Synthesis of Nitrogen-Doped Graphene via Thermal Annealing Graphite Oxide with Melamine and Its Excellent Electrocatalysis. *ACS Nano* **5(6)**, 4350-4358.
10. Yang, S., Feng, X., Wang, X., Muellen, K. (2011) Graphene-Based Carbon Nitride Nanosheets as Efficient Metal-Free Electrocatalysts for Oxygen Reduction Reactions. Angewandte Chemie, International Edition **50(23)**, 5339-5343, S5339/1-S5339/9.
11. Yang, S., Zhi, L., Tang, K., Feng, X., Maier, J., Muellen, K. (2012) Efficient Synthesis of Heteroatom (N or S)-Doped Graphene Based on Ultrathin Graphene Oxide-Porous Silica Sheets for Oxygen Reduction Reactions. *Advanced Functional Materials* **22(17)**, 3634-3640, S3634/1-S3634/6.

12. Wang, S., Zhang, L., Xia, Z., Roy, A., Chang, D. W., Baek, J.-B., Dai, L. (2012) BCN Graphene as Efficient Metal-Free Electrocatalyst for the Oxygen Reduction Reaction. *Angewandte Chemie, International Edition* **51(17)**, 4209-4212, S4209/1-S4209/7. (a) Wang, S., Iyyamperumal, E., Roy, A., Xue, Y., Yu, D., Dai, L. (2011) Vertically Aligned BCN Nanotubes as Efficient Metal-Free Electrocatalysts for the Oxygen Reduction Reaction: A Synergetic Effect by Co-Doping with Boron and Nitrogen. *AngewandteChemie, International Edition* **50**, 11756-11760.

13. Liang, J., Jiao, Y., Jaroniec, M., Qiao, S. Z. (2012) Sulfur and Nitrogen Dual-Doped Mesoporous Graphene Electrocatalyst for Oxygen Reduction with Synergistically Enhanced Performance. *Angewandte Chemie, International Edition* **51(46)**, 11496-11500.

14. (a) Yu, L., Pan, X., Cao, Xi., Hu, P., Bao, X. (2011) Oxygen reduction reaction mechanism on nitrogen-doped graphene: A density functional theory study. *Journal of Catalysis* **282(1)**, 183-190. (b) Fazio, G. Ferrighi, L., Di Valentin, C. (2014) Boron-doped graphene as active electrocatalysts for oxygen reduction reaction at a fuel-cell cathode. *Journal of Catalysis* **318**, 203-210.

15. Zheng, Y., Jiao, Y., Ge, L., Jaroniec, M., Qiao, S. Z. (2013) Two-Step Boron and Nitrogen Doping in Graphene for Enhanced Synergistic Catalysis. *Angewandte Chemie, International Edition* **52(11)**, 3110-3116.

16. Ma, Y., Sun, L., Huang, W., Zhang, L., Zhao, J., Fan, Q., Huang, W. (2011) Three-dimensional nitrogen-doped carbon nanotubes/graphene structure used as a metal-free electrocatalyst for the oxygen reduction reaction. *Journal of Physical Chemistry C* **115(50)**, 24592-24597.

17. Jeon, I.-Y., Choi, H.-J., Jung, S.-M., Seo, J.-M., Kim, M.-J., Dai, L., Baek, J.-B. (2013) Large-Scale Production of Edge-Selectively Functionalized Graphene Nanoplatelets via Ball Milling and Their Use as Metal-Free Electrocatalysts for Oxygen Reduction Reaction. *Journal of the American Chemical Society* **135(4)**, 1386-1393. (a) Jeon, I.-Y., Shin, Y. R., Sohn, G. J., Choi, H.-J., Bae, S. Y., Mahmood, J., Jung, S.-M., Seo, J.-M., Kim, M.-J., Chang, D. W., Dai, L., Baek, J.-B. (2012) *Proceedings of National Academy of Sciences U.S.A.* **109**, 5588-5593.

18. Wang, L., Ambrosi, A., Pumera, M. (2013) "Metal-Free" Catalytic Oxygen Reduction Reaction on Heteroatom-Doped Graphene is Caused by Trace Metal Impurities. *Angewandte Chemie, International Edition* **52(51)**, 13818-13821.

19. Wu, Z.-S., Ren, W., Wang, D.-W., Li, F., Liu, B., Cheng, H.-M. (2010) High-Energy MnO$_2$ Nanowire/Graphene and Graphene Asymmetric Electrochemical Capacitors, *ACS Nano* **4**, 5835–5842.

20. Wu, Z.-S., Yang, S., Sun, Y. Parvez, K., Feng, X., Müllen, K. (2012) 3D Nitrogen-Doped Graphene Aerogel-Supported Fe$_3$O$_4$ Nanoparticles as Efficient Electrocatalysts for the Oxygen Reduction Reaction, *Journal of American Chemical Society* **134**, 9082–9085.

21. Alammari, W., Govindhan, M., Chen, A. (2015) Modification of TiO2 Nanotubes with PtRu/Graphene Nanocomposites for Enhanced Oxygen Reduction Reaction. C*hemElectroChem* **2**, 2041–2047.

22. Tian, J., Ning, R., Liu, Q., Asiri, A.M., Al-Youbi, A. O., Sun, X. (2014) Three-Dimensional Porous Supramolecular Architecture from Ultrathin g-C3N4Nanosheets and Reduced Graphene Oxide: Solution Self-Assembly Construction and Application as a Highly Efficient Metal-Free Electrocatalyst for Oxygen Reduction Reaction, ACS Applied Materials & Interfaces **6**, 1011–1017.

23. Chen, P., Xiao, T,-Y., Li, S.-S., Yu, S.-H. (2013) A Nitrogen-Doped Graphene/Carbon Nanotube Nanocomposite with Synergistically Enhanced Electrochemical Activity. *Advanced Materials* **25**, 3192–3196.

24. Zhao, L., Sui, X.-L., Li, J.-L., Zhang, J.-J., Zhang, L.-M., Wang, Z.-B. (2016) 3D Hierarchical Pt-Nitrogen-Doped-Graphene-Carbonized Commercially Available Sponge as a Superior Electrocatalyst for Low-Temperature Fuel Cells, *ACS Applied Materials & Interfaces* **8**, 16026–16034.

25. Quesnel, E., Roux, F., Emieux, F., Faucherand, P., Kymakis, E., Volonakis, G., Giustino, F., Martín-García, B., Moreels, I., Gürsel, S. A., Yurtcan, A. B., Di Noto, V., Talyzin, A., Baburin, I., Tranca, D., Seifert, G., Crema, L., Speranza, G., Tozzini, V., Bondavalli, P., Pognon, G., Botas, C., Carriazo, D., Singh, G., Rojo, T., Kim, G., Yu, W., Grey, C. P., Pellegrini, V. (2015) *2D Materials* **2**, 030204.

26. Zhao, Y., Zhan, L., Tian, J., Nie, S., Ning, Z. (2011) Enhanced electrocatalytic oxidation of methanol on Pd/polypyrroleegraphene in alkaline medium. *Electrochimica Acta* **56,** 1967-1972.

27. Saito, A., Tsuji, H., Shimoyama, I., Shimizu, K., Nishina, Y. (2015) Highly durable carbon supported Pt catalysts prepared by hydrosilane-assisted nanoparticle deposition and surface functionalization., *Chemical Communications* **51,** 5883-5886.

28. Ren, F., Wang, H., Zhai, C., Zhu, M., Yue, R., Du, Y., Yang, P., Xu, J., Lu, W. (2014) Clean method for the synthesis of reduced graphene oxide-supported PtPd alloys with high electrocatalytic activity for ethanol oxidation in alkaline medium, *ACS Applied Materials Interfaces* **6** 3607-3614.

29. Habibi, B., Delnavaz, N. (2015) Pt-CeO$_2$/reduced graphene oxide nanocomposite for the electrooxidation of formic acid and formaldehyde, *RSC Advances* **5**, 73639-73650.

30. Zhou, Y., Hu, X.-C., Fan, Q., Wen, H.-R. (2016) Three-dimensional crumpled graphene as an electro-catalyst support for formic acid electro-oxidation, *Journal of Materials Chemistry A* **4**, 4587-4591.

31. Jiang, Y., Yan, Y., Chen, W., Khan, Y., Wu, J., Zhang, H., Yang, D. (2016) Single-crystalline Pd square nanoplates enclosed by {100} facets on reduced graphene oxide for formic acid electro-oxidation, *Chemical Communications* **52**, 14204-14207.

32. Zhang, K., Xiong, Z., Li, S., Yan, B., Wang, J., Du, Y. (2017) Cu$_3$P/RGO promoted Pd catalysts for alcohol electro-oxidation, *Journal of Alloys and Compounds* **706**, 89-96.

33. He, Z. Y., He, K., Robertson, A. W., Kirkland, A. I., Kim, D., Ihm, J., Yoon, E., Lee, G.-D., Warner, J. H. (2014) Atomic structure and dynamics of metal dopant pairs in graphene. *Nano Letters* **14**, 3766–3772.

34. Li, Y.W., Su, H. B., Chan, S. H., Sun, Q. (2016) CO$_2$ electroreduction performance of transition metal dimers supported on graphene: A theoretical study, *ACS Catalysis* **5**, 6658–6664.

35. Ruffieux, P., Wang, S. Y., Yang, B., Sánchez-Sánchez, C., Liu, J., Dienel, T., Talirz, L., Shinde, P., Pignedoli, C.A., Passerone, D. (2016) On-surface synthesis of graphene nanoribbons with zigzag edge topology, *Nature* **521**, 489–492.

36. Cai, J. M., Ruffieux, P., Jaafar, R., Bieri, M., Braun, T., Blankenburg, S., Muoth, M., Seitsonen, A.P., Saleh, M., Feng, X.L. (2010) Atomically precise bottom-up fabrication of graphene nanoribbons, *Nature* **466**, 470–473.

37. Kimouche, A., Ervasti, M.M., Drost, R., Halonen, S., Harju, A., Joensuu, P.M., Sainio, J., Liljeroth, P. (2015) Ultra-narrow metallic armchair graphene nanoribbons, *Nature Communications* **6**, 10177.

38. Zhu, G., Li, Y., Zhu, H., Su, H., Chan, S.H., Sun, Q. (2017) Enhanced CO$_2$ electroreduction on armchair graphene nanoribbons edge-decorated with copper, *Nano Research* **10(5)**, 1641–1650.

39. Allen, M.J., Tung, V.C., Kaner, R.B. (2009) Honeycomb carbon: a review of graphene, *Chemical Reviews* **110**, 132-145.

40. Kim, D., Ahmed, M.S.,  Jeon, S. (2012)  Different length linkages of graphene modified with metal nanoparticles for oxygen reduction in acidic media. *Journal of  Materials Chemistry*  **22**, 16353-16360.

41. Li, Y.M.,  Tang, L.H.,  Li, J.H. (2009) Preparation and electrochemical performance for methanol oxidation of Pt/graphene nanocomposites. *Electrochemistry Communications* **11**, 846-849.

42. Li, A.Z., Zhao, X., Hou, Y.N., Liu, H.J., Wu, L.Y., Qu, J.H. (2012) The electrocatalytic dechlorination of chloroacetic acids at electrodeposited Pd/Fe-modified carbon paper electrode. *Applied Catalysis B Environmental* **111-112**, 628-635.

43. Conner, W.C.,  Falconer, J.L. (1995) Spillover in heterogeneous catalysis. *Chemical Reviews*  **95**, 759-788.

**Future Reading**

[1] Ramaswamy, N., Mukerjee, S. (2011) Influence of Inner- and Outer-Sphere Electron Transfer Mechanisms during Electrocatalysis of Oxygen Reduction in Alkaline Media, *J. Phys. Chem. C* **115**, 18015-18026.

[2] Lv, R., Wang, H., Yu, H., Peng, F. (2017) Controllable Preparation of Holey Graphene and Electrocatalytic Performance for Oxygen Reduction Reaction, *Electrochim. Acta* **228,** 203-213.

[3] Farooque, M., Ghezel-Ayagh, H. (2003) *System design*, in *Handbook of Fuel Cells - Fundamentals, Technology and Applications,* Vielstich, V., Lamm, A., Gasteiger, H. A., (Eds.), Vol. 3, pp. 942-968, John Wiley & Sons, Chichester.

[4] Guo, S., Zhang, S., Sun, S. (2013) Tuning Nanoparticle Catalysis for the Oxygen Reduction Reaction, *Angew. Chem. Int. Ed.* **52**, 8526-8544.

[5] Huang, X., Qi, X., Boey, F., Zhang, H. (2012) Graphene-based composites, *Chem. Soc. Rev.* **41**, 666-686.

[6] Zoladek, S., Rutkowska, I. A., Blicharska, M., Miecznikowski, K., Ozimek, W., Orlowska, J., Negro, E., Di Noto, V., Kulesza, P. J. (2017) Evaluation of reduced-graphene-oxide-supported gold nanoparticles as catalytic system for electroreduction of oxygen in alkaline electrolyte, Electrochim. Acta **233,** 113-122.